\documentclass[epj-spec]{svjour}
\usepackage{graphics}
\usepackage{epsfig}
\usepackage{float,amsmath}
\usepackage{amssymb}
\begin{document}

\title{Suppression of forward dilepton production from an anisotropic quark-gluon plasma}
\author{
Mauricio Martinez\inst{1}\fnmsep\thanks{\email{guerrero@fias.uni-frankfurt.de}} 
\and 
Michael Strickland\inst{2,3}
}
\institute{Helmholtz Research School and Otto Stern School\\
  Goethe - Universit\"at Frankfurt am Main\\
  Ruth-Moufang-Str. 1,
  60438 Frankfurt am Main, Germany \and 
  Institute f\"ur Theoretische Physik and Frankfurt Institute for Advanced Studies \\
  Goethe - Universit\"at Frankfurt am Main\\
  Max-von-Laue-Stra\ss{}e 1,
  D-60438 Frankfurt am Main, Germany \and
  Department of Physics \\ 
  Gettysburg College \\
  Gettysburg, PA 17325
}
\abstract{
We calculate the rapidity dependence of leading-order medium dilepton 
yields resulting from a quark-gluon plasma which has a local 
time-dependent anisotropy in momentum space. We present a 
phenomenological model which includes temporal evolution of the plasma 
anisotropy parameter, $\xi$, and the hard momentum scale, $p_{\rm 
hard}$. Our model interpolates between 1+1 dimensional 
collisionally-broadened expansion at early times and 1+1 dimensional 
ideal hydrodynamic expansion at late times. Using our model, we find 
that at LHC energies, forward high-energy medium dilepton production 
would be suppressed by up to a factor of 3 if one assumes an 
isotropization/thermalization time of 2 fm/c. Therefore, it may be 
possible to use forward dilepton yields to experimentally determine 
the time of onset of locally isotropic hydrodynamic expansion of the 
quark-gluon plasma as produced in ultrarelativistic heavy-ion 
collisions. 
} %end of abstract

\maketitle
%

%%%%%%%%%%%%%%%%%%%%%%%%%%%%%%%%%%%%%%%%%%%%%%%%%%%%%%%%%%%%%%%%%%%%%%

\section{Introduction}
\label{intro}

Nucleus-nucleus collisions at high energies offer us the opportunity to 
study experimentally and theoretically a possible new state of matter 
formed by liberated partons known as a quark-gluon plasma (QGP). 
Nevertheless, the main goal is not only to discover this new state of 
matter but also to characterize experimentally many of its properties 
such as the thermodynamical aspects of this new phase. It is very 
important to know if the matter created after the collision is really 
thermalized. In this direction, an open question is to determine the 
isotropization and thermalization time of the matter, $\tau_{\rm iso}$ 
and $\tau_{\rm therm}$, respectively.\footnote{Hereafter, for 
simplicity, will assume that these two time scales are the same so 
that $\tau_{\rm therm}=\tau_{\rm iso}$.} Based on data from the 
Brookhaven National Lab's Relativistic Heavy Ion Collider (RHIC) early 
studies found that for $p_T \lesssim 2$ GeV, the elliptic flow, $v_2$, 
was very well described by models which assumed ideal hydrodynamic 
behavior starting at very early times $\tau \sim$ $0.6$ 
fm/c~\cite{Huovinen:2001cy,Hirano:2002ds,Tannenbaum:2006ch}. These 
fits of $v_2$ indicated that the matter could be modeled as a 
nearly-perfect fluid and hence implied fast thermalization of the 
matter created at RHIC energies. However, recent results from 
conformal relativistic viscous hydrodynamics~\cite{Luzum:2008cw} have 
shown that these initial estimates for the 
isotropization/thermalization time of the plasma are not completely 
reliable due to poor knowledge of the proper initial conditions (CGC 
versus Glauber), details of plasma hadronization such as the choice of 
the proper freezeout time and the subsequent hadronic cascade, etc. 
Now it seems that isotropization times up to $\tau_{\rm iso} \sim$ 2 
fm/c are not ruled out by RHIC data. Therefore, additional theoretical 
and experimental input are necessary to further constrain this time.

One way to increase our understanding of the pre-equilibrium stage of 
QGP evolution is by studying independent observables which are 
sensitive to early-time dynamics. One good candidate is high-energy 
dileptons.\footnote{We mean by high-energy dileptons, lepton pairs 
with pair transverse momentum ($p_T$) or invariant mass ($M$) greater 
than 1 GeV.} Due to their large mean free path, lepton pairs can leave 
the strongly interacting region carrying information about the 
earliest times after the nuclear collision. In 
this work, we calculate the rapidity dependence of dilepton pair 
production from a QGP which has a time-dependent anisotropy in 
momentum space.

Most previous phenomenological treatments of dilepton production have 
assumed that the plasma is thermalized rapidly with $\tau_{\rm iso}$ 
on the same order as the parton formation time, $\tau_0$. However, due 
to the rapid 1+1 longitudinal expansion of the plasma at early-times, 
this assumption seems to be rather strong because it ignores the 
momentum-space anisotropy developed along the beam axis. Anisotropies 
in momentum-space are intimately related with Weibel instabilities. 
Recently, it has been shown that the presence of chromomagnetic Weibel 
instabilities developed at early times of the collision may affect the 
value of $\tau_{\rm iso}$ 
\cite{Romatschke:2003ms,Mrowczynski:2000ed,Arnold:2003rq,% 
Romatschke:2004jh,Rebhan:2008uj}. However, it is still not clear how 
much $\tau_{\rm iso}$ is affected by the presence of these 
instabilities at phenomenologically accessible collision energies.

Due to the lack of a precise physical picture of the pre-equilibrium 
QGP, one can instead implement simple models which capture the essence 
of the complex dynamics at early-times of the collision. This will be 
our approach in the present work. We propose a simple phenomenological 
model for the time-dependence of the hard momentum scale, $p_{\rm 
hard}$, and the plasma anisotropy parameter, $\xi=\langle 
p_T^2\rangle/2 \langle p_L^2\rangle - 1$. The new aspect in this work 
compared to our recent proposals for the pre-equilibrium stage of the 
QGP \cite{Mauricio:2007vz,Martinez:2008rr,Martinez:2008di}, is the 
inclusion of the rapidity dependence of the quark and anti-quark 
distribution functions.  We use fits to experimental data which are 
available from AGS through RHIC data to constrain the rapidity 
dependence of the parton distribution functions.  Since this rapidity 
dependence is not perfectly flat we implicitly include effects of the 
breaking of longitudinal boost invariance of the system.

To make our final phenomenological predictions for dilepton yields, we 
introduce three different parameters: (1) the parton formation time, 
$\tau_0$, which is time at which the partons created during the 
initial hard collision of the nuclei become decoherent and 
approximately on-shell;\footnote{This time can be estimated from the 
nuclear saturation scale, i.e., $\tau_0\sim Q_s^{-1}$. For LHC 
energies, $Q_s\simeq$ 2 GeV implying $\tau_0\simeq$ 0.1 fm/c} (2) 
$\tau_{\rm iso}$ which is the proper time when the system starts to 
undergo ideal hydrodynamical expansion; and (3) $\gamma$ which sets 
the sharpness of the transition to hydrodynamical behaviour. We assume 
that for times of the order of $\tau_0$ but smaller than $\tau_{\rm 
iso}$, the system expands like a collisionally-broadened plasma and 
smoothly changes to hydrodynamical expansion for times long compared 
with $\tau_{\rm iso}$. In this work we extend previous calculations 
\cite{Mauricio:2007vz,Martinez:2008rr,Martinez:2008di} considering the 
impact of the momentum-space anisotropies on the full rapidity 
dependence of medium dilepton production. We find that at LHC 
energies, dilepton yields are suppressed by a factor of 3 around 
$y\sim$ 9 if one chooses $\tau_{\rm iso}$=2 fm/c.

This manuscript is organized as follows: In Sec. \ref{sec:1} we 
calculate the dilepton production rate at leading order using an 
anisotropic phase space distribution. In Sec. \ref{sec:2} we present 
the interpolating model from collisionally-broadened to ideal 
hydrodynamical expansion including the rapidity dependence of the 
quark and anti-quark distribution functions. In Sec. \ref{sec:5} we 
present dilepton yields as a function of the rapidity for different 
values of $\tau_{\rm iso}$. Finally, we present our conclusions and 
give an outlook in the Sec. \ref{sec:6}.

%%%%%%%%%%%%%%%%%%%%%%%%%%%%%%%%%%%%%%%%%%%%%%%%%%%%%%%%%%%%%%%%%%%%%%%%%%%%%%%%%%%%%%%%%%%%%%%%%%%%%%%%%%%%%%%%%%%%%%%%%%%%%%%%%%%%%%%%%%%%%%%%%%%%%%%%%%%%%%%%%%%%%%%%%%%%%%%%%%%%%%%%%%%%%%%%%%%%%%%%%%%%%%%%%%%%%%%%%%

\section{Dilepton rate}
\label{sec:1}

From relativistic kinetic theory, the dilepton production rate 
$dN^{l^+l^-}/d^4Xd^4P\equiv d R^{l^+l^-}/d^4P$ (i.e. the number of 
dileptons produced per space-time volume and four dimensional 
momentum-space volume) at leading order in the electromagnetic 
coupling, $\alpha$, is given by 
\cite{Kapusta:1992uy,Dumitru:1993vz,Strickland:1994rf}:
\begin{equation}
\frac{d R^{l^+l^-}}{d^4P} = \int \frac{d^3{\bf p}_1}{(2\pi)^3}\,\frac{d^3{\bf p}_2}{(2\pi)^3}
			\,f_q({\bf p}_1)\,f_{\bar{q}}({\bf p}_2)\, \it{v}_{q\bar{q}}\,\sigma^{l^+l^-}_{q\bar{q}}\,
			\delta^{(4)}(P-p_1-p_2)
      \; ,
\label{eq:annihilation1}
\end{equation}
where $f_{q,{\bar q}}$ is the phase space distribution function of the 
medium quarks (anti-quarks), $\it{v}_{q\bar{q}}$ is the relative 
velocity between quark and anti-quark and $\sigma^{l^+l^-}_{q\bar{q}}$ 
is the total cross section. Since we will be considering high-energy 
dilepton pairs with center-of-mass energies much greater than the 
dilepton mass we can ignore the finite dilepton mass corrections and use 
simply $\sigma^{l^+l^-}_{q\bar{q}} = 4 \pi \alpha^2 / 3 M^2$. In 
addition, we assume that the distribution function of quarks and 
anti-quarks is the same, $f_{\bar q}=f_q$.

Following Ref.~\cite{Romatschke:2003ms}, we will consider the case 
that the anisotropic quark/anti-quark phase-space distributions are 
azimuthally symmetric in momentum space about a direction specified by 
${\bf\hat{n}}$ and can be obtained from an arbitrary isotropic 
phase space distribution by squeezing ($\xi>0$) or stretching 
($\xi<0$) this isotropic distribution function along 
${\bf\hat{n}}$, i.e.,
\begin{equation}
f_{q,{\bar q}}({\bf p},\xi,p_{\rm hard})=f^{iso}_{q,{\bar 
q}}(\sqrt{{\bf p^2}+\xi({\bf p\cdot \hat{n}}){\bf^2}},p_{\rm hard}) \; ,
\label{eq:distansatz}
\end{equation}
where $p_{\rm hard}$ is the hard momentum scale, $\hat{\bf n}$ is the 
direction of the anisotropy and $\xi >0$ is a parameter that reflects 
the strength and type of anisotropy. In the isotropic case, $p_{\rm 
hard}$ is identified with the temperature of the system and $\xi\equiv 
0$. Using the anisotropic distribution given by Eq.~(\ref{eq:distansatz}) 
into Eq.~(\ref{eq:annihilation1}) we obtain:\footnote{Details of the 
calculation are presented in Ref.~\cite{Martinez:2008di}.}
\begin{eqnarray}
\frac{d R^{l^+l^-}}{d^4P}&=&
\frac{5\alpha^2}{18\pi^5}\int_{-1}^1d(\cos\theta_{p_1})
\int_{a_+}^{a_-}\frac{dp_1}{\sqrt{\chi}}\,p_1\hspace{0.1cm}f_q\left({\sqrt{\bf p_1^2(1+\xi(\tau)\cos^2\theta_{p_1})}},p_{\rm hard}(\tau,\eta)\right)\nonumber \\
&\times & f_{\bar{q}}\left(\sqrt{{\bf(E-p_1)^2+\xi(\tau)(p_1\cos\theta_{p_1}-P\cos\theta_P)^2}},p_{\rm hard}(\tau,\eta)\right),
\label{scattering}
\end{eqnarray}
with
\begin{eqnarray*}
\chi&=&\,4\,P^2\,p_1^2\,\sin^2\theta_P\,\sin^2\theta_{p_1}-(2p_1(E-P\cos\theta_P\cos\theta_{p_1})-M^2)^2 \; , \\
a_{\pm}&=&\frac{M^2}{2(E-P\cos (\theta_P\pm\theta_{p_1}))} \; .
\end{eqnarray*}

For phenomenological predictions of expected dilepton yields, we model 
the space-time dependence of $p_{\rm hard}$ and $\xi$.  In a realistic 
non-boost-invariant system $p_{\rm hard}$ will depend of the rapidity 
on the quark and anti-quark and hence the rate itself will depend not 
only on the difference of $y-\eta$ but also on $\eta$ itself.  In the 
next sections we will give details of how this dependence is 
introduced.  Once obtained we can calculate the final rapidity spectra 
by integrating over phase space:
\begin{equation}
  \frac{dN^{l^+l^-}}{dy}=\pi R^2_T\int dM^2 d^2p_T\int_{\tau_0}^{\tau_f}\int_{-\infty}^{\infty}\frac{dR^{l^+l^-}}{d^4P}\tau d\tau d\eta\hspace{0.2cm}, \label{yspectrum}
\end{equation}
where $R_T\,=\,1.2\,A^{1/3}$ fm is the radius of the nucleus in the 
transverse plane. This expression is evaluated in the center-of-mass 
(CM) frame while the differential dilepton rate is calculated in the 
local rest frame (LR) of the emitting region. Then, the dilepton pair 
energy has to be understood as $E_{LR}=p_T\,\cosh\,(y-\eta)$ in the 
differential dilepton rate $dR^{l^+l^-}/d^4P$. Substituting 
Eq.~(\ref{scattering}) into Eq.~(\ref{yspectrum}), we obtain the 
dilepton spectrum as a function of the rapidity including the effect 
of a time-dependent momentum anisotropy.

One can be worried if either transverse expansion or mixed/hadronic 
phases in Eq.~(\ref{yspectrum}) will affect the production of 
high-energy dileptons presented here. Fortunately, in the kinematic 
region studied here this is not the case and these effects turn out to 
be negligible (1-2\% effect) compared with the longitudinal expansion 
\cite{Mauricio:2007vz}.

%%%%%%%%%%%%%%%%%%%%%%%%%%%%%%%%%%%%%%%%%%%%%%%%%%%%%%%%%%%%%%%%%%%%%%

\section{Space Time-Model}
\label{sec:2}

In this section, we extend previous models proposed for the 
pre-equilibrium stage of the QGP 
\cite{Mauricio:2007vz,Martinez:2008rr,Martinez:2008di} by including a phenomenologically 
realistic rapidity dependence of the parton distributions. Before 
presenting the details of the model, we remind the reader of some 
general relations.

Firstly, the plasma anisotropy parameter is related with the average 
longitudinal and transverse momentum of the hard particles through the 
relation:
\begin{equation}
\label{anisoparam}
 \xi=\frac{\langle p_T^2\rangle}{2\langle p_L^2\rangle} - 1 \; .
\end{equation}

We can obtain two limiting cases from the last expression. For an 
isotropic plasma, we have that $\xi$=0, i.e., $\langle p_T^2\rangle = 
2\langle p_L^2\rangle$. Another possibility is that the partons initially
undergo 1d free streaming 
expansion, where the partons expand freely along the longitudinal 
axis. Using the free streaming distribution function, it is possible 
to show that the transverse and longitudinal momentum scales as 
\cite{Rebhan:2008uj,Martinez:2008di,Baier:2000sb}:
\begin{subequations}
 \begin{align}
  \langle p_T^2 \rangle_{\rm f.s.} &\propto 2 \, T_0^2 \; ,\\
  \langle p_L^2 \rangle_{\rm f.s.} &\propto T_0^2 \frac{\tau_0^2}{\tau^2} \; .
 \end{align}
\end{subequations}
Inserting these expressions into Eq.~(\ref{anisoparam}), one obtains 
$\xi_{f.s.}(\tau) = \tau^2/\tau_0^2-1$.  The free streaming result for 
$\xi$ is the upper-bound on possible momentum-space anisotropies 
developed during 1d expansion by causality. Modifications to the 
Eq.~(\ref{anisoparam}) result from the different kinds of interactions 
between the partons, as it is discussed below. In this work, for 
simplicity, we will not study explicitly the possibility of 1d free 
streaming expansion since, in reality, this is a rather extreme 
assumption which requires that the partons do not interact at all. 
Below we will give details of a more realistic model of the time 
evolution of the average transverse and longitudinal pressures. 
However, note that the relation given in Eq.~(\ref{anisoparam}) is 
completely general and can be applied in all cases.

Secondly, for a given anisotropic phase space distribution of the 
form specified in Eq.~(\ref{eq:distansatz}), the local energy density 
can be factorized as:
\begin{eqnarray}
\label{energy}
{\cal E}(p_{\rm hard},\xi) &=& \int \frac{d^3{\bf p}}{(2\pi)^3}\hspace{0.1cm}p\hspace{0.1cm}
	f_{\rm iso}(\sqrt{{\bf p^2}+\xi({\bf p\cdot \hat{n}}{\bf^2})},p_{\rm hard}) \; , \\ 
&=& {\cal E}_0(p_{\rm hard}) \, {\cal R}(\xi) \; ,\nonumber
\end{eqnarray}
where ${\cal E}_0$ is the initial local energy density deposited in 
the medium at $\tau_0$ and
\begin{equation}
{\cal R}(\xi) \equiv \frac{1}{2}\Biggl(\frac{1}{1+\xi}+\frac{\arctan\sqrt{\xi}}{\sqrt{\xi}} \Biggr) \; .
\label{calRdef}
\end{equation}
We mention that $p_{\rm hard}$ depends implicitly on proper time and 
space-time rapidity and that $\lim_{\xi \rightarrow 0} {\cal R}(\xi) = 
1$ and $\lim_{\xi \rightarrow \infty} {\cal R}(\xi) = 1/\sqrt{\xi}$.

\subsection{Momentum-space broadening and plasma instabilities effect}
\label{sec:3}

The ratio between the average longitudinal and transverse momentum 
needed to compute $\xi$ using Eq.~(\ref{anisoparam}) is modified from 
the free streaming case if collisions between the partons are taken 
into account. To include all relevant $N \leftrightarrow N$ Boltzmann 
collision terms plus mean field interactions (Vlasov term) is a very 
difficult task. As a first approximation, one can consider the effect 
of elastic collisions in the broadening of the longitudinal momentum 
of the particles. This was the approach in the original version of the 
bottom-up scenario \cite{Baier:2000sb}. In the first stage of this 
scenario, $1\ll Q_s\tau\ll\alpha_s^{3/2}$, initial hard gluons have 
typical momentum of order $Q_s$ and occupation number of order 
$1/\alpha_s$. Due to the fact that the system is expanding at the 
speed of light in the longitudinal direction, the density of hard 
gluons decreases with time, $N_g \sim Q_s^3/(\alpha_s Q_s\tau)$. If 
there were no interactions this expansion would be equivalent to 1+1 
free streaming and the longitudinal momentum $p_L$ would scale like 
$1/\tau$. However, once elastic $2\leftrightarrow 2$ collisions of 
hard gluons are taken into account, the ratio between the longitudinal 
momentum $p_L$ and the typical transverse momentum of a hard particle 
$p_T$ decreases as \cite{Baier:2000sb}:
\begin{equation}
\label{ptbroadbottom}
\frac{\langle p_L^2 \rangle}{\langle p_T^2 \rangle} \propto (Q_s\tau)^{-2/3} \; .
\end{equation}
Assuming, as before, isotropy at the formation time, 
$\tau_0=Q_s^{-1}$, this implies that for a collisionally-broadened
expansion, $\xi (\tau)=(\tau/\tau_0)^{2/3}-1$.

Note that, as obtained in Ref.~\cite{Baier:2000sb}, when collisions 
are included, it is implicitly assumed that the elastic cross-section 
is screened at long distances by an isotropic real-valued Debye mass. 
This is not guaranteed in an anisotropic plasma since the Debye mass 
can be become complex due to the chromo-Weibel instability. In 
general, any anisotropic distribution causes the presence of negative 
eigenvalues in the self-energy of soft gluons 
\cite{Romatschke:2003ms,Mrowczynski:2000ed,Arnold:2003rq,% 
Romatschke:2004jh,Rebhan:2008uj}. Such negative eigenvalues indicate 
instabilities, which result in exponential growth of chromo-electric 
and magnetic fields, $E^a$ and $B^a$, respectively. These fields deflect 
the particles and how much deflection occurs will depend on the amplitude 
and domain size of the induced chromofields. Currently, the precise 
parametric relation between the plasma anisotropy parameter and the 
amplitude and domain size of the chromofields is not known from first 
principles.

If one would like to include the momentum-space broadening and the 
impact of the plasma instabilities, phenomenologically this can be 
achieved by generalizing the temporal dependence of $\xi (\tau)$ to:
\begin{equation}
\xi(\tau,\delta) = \left( \frac{\tau}{\tau_0} \right)^\delta - 1 \; .
\label{broadenedxi}
\end{equation}
The exponent $\delta$ contains the physical information about the 
particular type of momentum-space broadening which occurs due to plasma
interactions. Two limiting cases for this exponent are 
the ideal hydrodynamic and free streaming expansion. In the 1+1 
hydrodynamical limit, $\delta \equiv 0$ and then $\xi \rightarrow 0$. 
For  $\delta \equiv 2$, one recovers the 1+1 dimensional free 
streaming case, $\xi \rightarrow \xi_{\rm f.s.} = (\tau /\tau_0)^2-1$. 
For general $\delta$ between these limits one obtains the proper-time 
dependence of the energy density and temperature by substituting 
(\ref{broadenedxi}) into the general expression for the factorized 
energy density (\ref{energy}) to obtain ${\cal E}(\tau,\delta) = {\cal 
E}_0(p_{\rm hard}) \, {\cal R}(\xi(\tau,\delta))$. The inclusion of the 
rapidity dependence of the parton distribution functions is discussed in 
Sec.~\ref{sec:4}.

Different values of $\delta$ arise dynamically from the different 
processes contributing to parton isotropization. For example, elastic
collisional-broadening results in Eq.~(\ref{ptbroadbottom}) and hence 
$\delta=2/3$.  Recently, some authors have considered the values of 
$\delta$ resulting from processes associated with the chromo-Weibel 
instability presented at the earliest times after the initial nuclear 
impact \cite{Bodeker:2005nv,Arnold:2005qs,Arnold:2007cg}:
\begin{equation}
\label{ptbroadproposals1}
\frac{\langle p_L^2 \rangle}{\langle p_T^2 \rangle}\sim(Q_s\tau)^{-\frac{1}{2}\bigl(\frac{1}{1+\nu}\bigl)} \; ,
\end{equation}
where
\begin{equation}
 \label{ptbroadproposals2}
\nu=\left\{ \begin{aligned}
0 \hspace{0.2cm}&\text{Ref.\cite{Bodeker:2005nv} \; ,}\\
1 \hspace{0.2cm}&\text{Ref.\cite{Arnold:2005qs} \; ,}\\
2 \hspace{0.2cm}&\text{Nielsen-Olesen limit, Ref.\cite{Arnold:2007cg}} \;.
          \end{aligned}
	  \right.
\end{equation}
Summarizing, the coefficient $\delta$ in Eq.~(\ref{broadenedxi}) takes 
on the following values
\begin{equation} 
\label{anisomodels}
\delta=\left\{ 
\begin{aligned}
2 \hspace{0.2cm}&\text{Free streaming expansion} \; ,\\
2/3 \hspace{0.2cm}&\text{Collisional-Broadening, Ref.\cite{Baier:2000sb}\; ,}\\
1/2 \hspace{0.2cm}&\text{Ref.\cite{Bodeker:2005nv}}\; , \\
1/4 \hspace{0.2cm}&\text{Ref.\cite{Arnold:2005qs}\; ,}\\
1/6 \hspace{0.2cm}&\text{Nielsen-Olesen limit, Ref.\cite{Arnold:2007cg}}\; ,\\
0\hspace{0.2cm}&\text{Hydrodynamic expansion} \; .
\end{aligned}
\right.
\end{equation}
%

%%%%%%%%%%%%%%%%%%%%%%%%%%%%%%%%%%%%%%%%%%%%%%%%%%%%%%%%%%%%%%%%%%%%%%%%%%%%%%
\begin{figure*}[t]
\begin{center}
 \includegraphics[width=10cm]{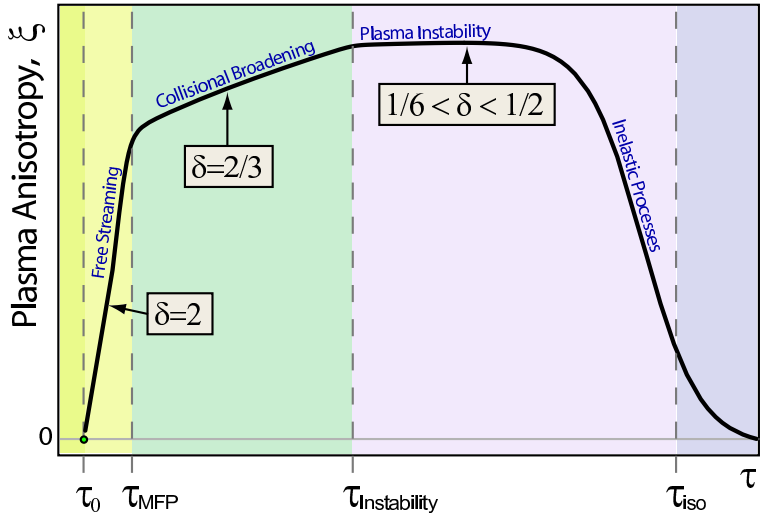}
\end{center}
\vspace{-2mm}
\caption{Sketch of the time dependence the plasma anisotropy indicating the 
various time scales and processes taking place.  Here $\tau_{\rm MFP}$ 
is the mean time between elastic collisions (mean-free time) and $\tau_{\rm 
Instability}$ is the time at which plasma-instability induced soft 
modes have grown large enough to affect hard particle dynamics.}
\label{fig:1}
\end{figure*}
%%%%%%%%%%%%%%%%%%%%%%%%%%%%%%%%%%%%%%%%%%%%%%%%%%%%%%%%%%%%%%%%%%%%%%%%%%%%%%

In Fig.~\ref{fig:1} we sketch the time-dependence of the plasma 
anisotropy parameter indicating the time scales at which the various 
processes become important.  At times shorter than the mean time 
between successive elastic scatterings, $\tau_{\rm MFP}$, the system 
will undergo 1+1 dimensional free streaming with $\delta=2$. For times 
long compared to $\tau_{\rm MFP}$ but short compared to $\tau_{\rm 
Instability}$ the plasma anisotropy will grow with the 
collisionally-broadened exponent of $\delta=2/3$. Here $\tau_{\rm 
Instability}$ is the time at which instability-induced soft gauge 
fields begin to influence the hard-particles' motion. When $\tau_{\rm 
Instability} < \tau < \tau_{\rm iso}$ the plasma anisotropy grows with 
the slower exponent of $\delta = 1/6 \ldots 1/2$ due to the bending of 
particle trajectories in the induced soft-field background.  At times 
large compared to $\tau_{\rm Instability}$ inelastic processes are 
expected to drive the system back to isotropy \cite{Baier:2000sb}. We 
note here that for small $\xi$ and realistic couplings it has been 
shown \cite{Schenke:2006xu} that one cannot ignore the effect of 
collisional-broadening of the distribution functions and that this may 
completely eliminate unstable modes from the spectrum.

Based on such a sketch, one could try to construct a detailed model 
which includes all of the various time scales and study the dependence 
of the process under consideration on each.  However, there are 
theoretical uncertainties in each of these time scales and their 
dependences on experimental conditions. We choose to use a simpler 
approach in which we will construct a phenomenological model which 
smoothly interpolates the coefficient $\delta$ from the 1d 
collisionally-broadened expansion to 1d hydrodynamical expansion, 
i.e., $2/3\geq \delta \geq 0$.

In the model we introduce a transition width, $\gamma^{-1}$, which 
governs the smoothness of the transition from the initial value of 
$\delta =2/3$ to $\delta=0$ at $\tau \sim \tau_{\rm iso}$. The 
collisionally-broadened interpolating model provides us a realistic 
estimate of the effect of plasma anisotropies. Note that by using such 
a smooth interpolation one can achieve a reasonable phenomenological 
description of the transition from non-equilibrium to equilibrium 
dynamics which should hopefully capture the essence of the physics. In 
the next section we will give mathematical definitions for the model. 

%%%%%%%%%%%%%%%%%%%%%%%%%%%%%%%%%%%%%%%%%%%%%%%%%%%%%%%%%%%%%%%%%%%%%%%%%%%%%%%%%%%%%%%%%%%%%%%%%%%%%%%%%%%%%%%%%%%%%%%%%%%%%%%%%%%%%%%%%%%%%%%%%%%%%%%%%%%%%%%%%%%%%%%%%%%%%%%%%%%%%%%%%%%%%%%%%%%%%%%%%%%%%%%%%%%%%%%%%%

\subsection{Interpolating model for collisionally broadened expansion}
\label{sec:4}

In order to construct an interpolating model between 
collisionally-broadened and hydrodynamical expansion, we introduce 
the smeared step function:
\begin{equation}
\lambda(\tau,\tau_{\rm iso},\gamma) \equiv \frac{1}{2} \left({\rm 
tanh}\left[\frac{\gamma (\tau-\tau_{\rm iso})}{\tau_{\rm iso}} \right]+1\right) \; ,
\end{equation}
where $\gamma^{-1}$ sets the width of the transition between 
non-equilibrium and hydrodynamical evolution in units of $\tau_{\rm 
iso}$. In the limit when $\tau \ll \tau_{\rm iso}$, we have $\lambda 
\rightarrow 0$ and when $\tau \gg \tau_{\rm iso}$ we have $\lambda 
\rightarrow 1$. Physically, the energy density ${\cal E}$ should be 
continuous as we change from the initial non-equilibrium value of 
$\delta$ to the final isotropic $\delta=0$ value appropriate for ideal 
hydrodynamic expansion.  Once the energy density is specified, 
this gives us the time dependence of the hard momentum scale. 
We find that for general $\delta$ this can be accomplished with the 
following model:
\begin{subequations}
\label{eq:modelEQs}
\begin{align}
\xi(\tau,\delta) &= \left(\tau/\tau_0\right)^{\delta(1-\lambda(\tau))} - 1 \; , \label{xidependence}\\
{\cal E}(\tau, \eta) &= {\cal E}_0 \; {\cal R}\left(\xi\right) \; \bar{\cal U}^{4/3}(\tau) \; F^4(\eta) \; ,\\
p_{\rm hard}(\tau, \eta) &= T_0 \; \bar{\cal U}^{1/3}(\tau)\;F(\eta) \; ,
\label{pdependence}
\end{align}
\end{subequations}
with ${\cal R}(\xi)$ defined in Eq.~(\ref{calRdef}) and for fixed final multiplicity we have:
\begin{subequations}
\begin{eqnarray}
{\cal U}(\tau) &\equiv& \left[{\cal R}\!\left(\left(\tau_{\rm iso}/\tau_0\right)^\delta- 
1\right)\right]^{3\lambda(\tau)/4} \left(\frac{\tau_{\rm iso}}{\tau}\right)^{1 - \delta\left(1-\lambda(\tau)\right)/2} \; , \nonumber \\
\bar{\cal U}(\tau) &\equiv& {\cal U}(\tau) \, / \, {\cal U}(\tau_{\rm iso}^+) \; , \\
{\cal U}(\tau_{\rm iso}^+) &\equiv& \lim_{\tau\rightarrow\tau_{\rm iso}^+} {\cal U}(\tau) 
  = \left[{\cal R}\!\left(\left(\tau_{\rm iso}/\tau_0\right)^\delta-1\right)\right]^{3/4} 
	  \left(\frac{\tau_{\rm iso}}{\tau_0}\right) \; .
\end{eqnarray}
\label{Udeff}
\end{subequations}
and $\delta = 2/3$ for the case of 1d collisionally broadened expansion
interpolating to 1d ideal hydrodynamic expansion.

Recently, models for the pre-equilibrium stage in the presence of 
momentum-space anisotropies have been applied in the central rapidity 
region of high-energy dileptons 
\cite{Mauricio:2007vz,Martinez:2008rr,Martinez:2008di}. Here we extend 
these previous analyses through the inclusion of rapidity dependence 
of the parton hard momentum scale by using a phenomenologically 
constrained ``rapidity profile function,'' $F(\eta)$. Our main goal 
with this modification is to explore the phenomenological consequences 
in the forward rapidity region where the effect of early-time 
anisotropies are expected to be maximal. We are not attempting to 
describe the physics of the forward rapidity region from first 
principles,\footnote{Some proposals have been mentioned in the 
literature, see 
Ref.~\cite{Renk:2004yv,Hirano:2004en,Hirano:2001eu,Morita:2002av}} 
instead, we implement a Gaussian fit profile for the rapidity 
dependence which successfully describes experimentally observed pion 
rapidity spectra from AGS to RHIC energies 
\cite{Bearden:2004yx,Park:2001gm,Back:2005hs,Veres:2008nq,Bleicher:2005tb} and use this to 
extrapolate to LHC energies:
\begin{equation}
\label{yprofile}
 F(\eta)=\exp \Biggl(-\frac{\eta^2}{2\sigma_\eta^2}\Biggr) \; ,
\end{equation}
with
\begin{equation}
\label{width}
 \sigma_\eta^2=\frac{8}{3}\frac{c_s^2}{(1-c_s^4)}\ln \left(\sqrt{s_{NN}}/2 m_p\right) \; , 
\end{equation}
where $c_s$ is the sound velocity and $m_p$ is the proton mass.

Note that once the rapidity dependence in the parton momentum 
distribution functions is implemented boost invariance along the 
longitudinal axis breaks down. This procedure leads to a violation of 
the conservation laws expressed by hydrodynamics unless a finite 
baryon chemical potential is introduced~\cite{Dumitru:1995bu}. It is 
possible to demonstrate that for a longitudinal scaling expansion, 
$\partial {\cal P}(T,\mu)/\partial\eta$=0, where ${\cal P}$ is the 
pressure and $T$ is the temperature. This 
condition is equivalent to:\footnote{Partial derivatives with respect 
to $\eta$ are performed at constant $\tau$.}
\begin{equation}
 s\frac{\partial T}{\partial \eta}+ n\frac{\partial \mu}{\partial \eta}=0 \; ,
\end{equation}
where $s$ and $n$ denote the entropy density and particle number 
density, respectively. In the present context don't have to worry 
about the presence of finite chemical potentials since we are 
considering high-energy dilepton production and  $E/T\gg\mu/T$ is 
satisfied. Therefore, in the differential dilepton rate $dR/d^4P$, 
Eq.(\ref{eq:annihilation1}), the product of the distribution functions 
$f_q({\bf p}_q,+\mu)\;f_{{\bar q}}({\bf p}_{{\bar q}},-\mu)\approx 
f_q({\bf p}_q)\;f_{{\bar q}}({\bf p}_{{\bar q}})$.

\begin{figure*}[t]
\begin{center}
 \includegraphics[width=8cm]{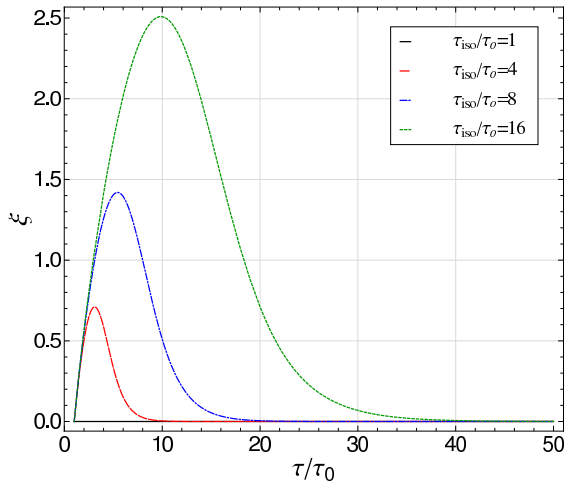}
\end{center}
\vspace{-2mm}
\caption{Temporal evolution of the plasma anisotropy parameter using our
collisionally-broadened interpolating model for four different isotropization
times $\tau_{\rm iso} \in 
\{1,4,6,18\} \, \tau_0$. The transition width is taken to be $\gamma = 
2$.  To convert to physical scales use $\tau_0 \sim 0.1$ fm/c for LHC 
energies.}
\label{fig:2}
\end{figure*}
%%%%%%%%%%%%%%%%%%%%%%%%%%%%%%%%%%%%%%%%%%%%%%%%%%%%%%%%%%%%%%%%%%%%%%

%%%%%%%%%%%%%%%%%%%%%%%%%%%%%%%%%%%%%%%%%%%%%%%%%%%%%%%%%%%%%%%%%%%%%%
\begin{figure*}[t] \begin{center} 
\includegraphics[width=14.3cm]{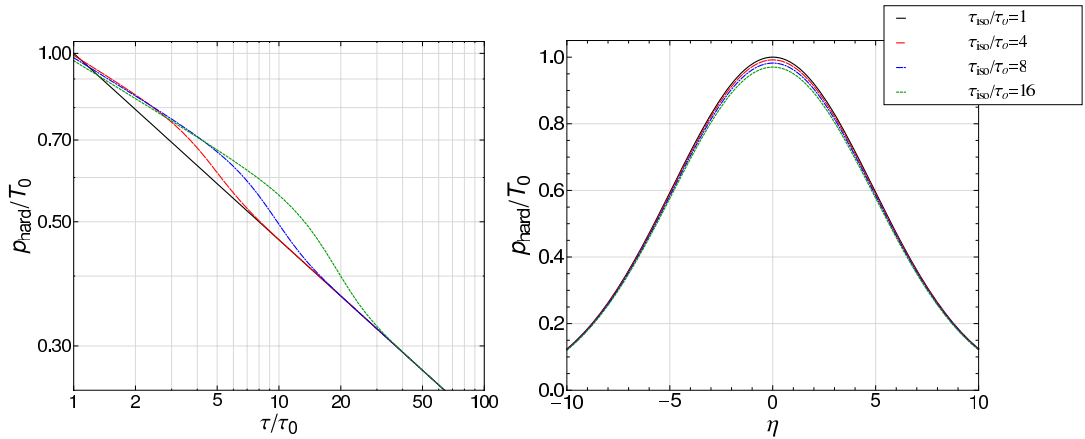} 
\end{center} 
\vspace{-2mm} 
\caption{Temporal evolution (left panel) and rapidity dependence 
(right panel) of the hard momentum scale, $p_{\rm hard}$ using our 
fixed final multiplicity collisionally-broadened interpolating model 
for the hard momentum scale for four different isotropization times 
$\tau_{\rm iso} \in \{1,4,8,16\} \, \tau_0$. The transition width is 
taken to be $\gamma$ = 2. To convert to physical units use 
$\tau_0\sim$ 0.1 fm/c for LHC energies. For the rapidity dependence of 
$p_{\rm hard}$ (right panel) we used a constant value of $\tau\sim$ 
0.1 fm/c and the width $\sigma_\eta^2 \sim$ 8. } 
\label{fig:3} 
\end{figure*} 
%%%%%%%%%%%%%%%%%%%%%%%%%%%%%%%%%%%%%%%%%%%%%%%%%%%%%%%%%%%%%%%%%%%%%%

In Fig.~\ref{fig:2}, the temporal evolution of the anisotropy 
parameter $\xi(\tau)$ is plotted using Eq.~(\ref{xidependence}). In 
Fig.~\ref{fig:3}, we show the time and rapidity dependence of $p_{\rm 
hard}(\tau ,\eta)$ (right and left panel, respectively) using 
Eq.~(\ref{pdependence}).

%%%%%%%%%%%%%%%%%%%%%%%%%%%%%%%%%%%%%%%%%%%%%%%%%%%%%%%%%%%%%%%%%%%%%%

\section{Results}
\label{sec:5}

In this section, we will present our predicted dilepton yields as a 
function of the rapidity from a Pb-Pb collision at LHC full beam 
energy, $\sqrt{s_{NN}}$= 5.5 TeV. At this center-of-mass energy we use 
$\tau_0$= 0.088 fm/c, $T_0$= 845 MeV, $R_T$= 7.1 fm and the critical 
temperature $T_c$=160 MeV. The kinematic cuts in the transverse 
momentum and invariant mass of the dilepton yields are indicated in 
the corresponding results. Also, we use $c_s^2=1/3$ and $m_p=0.938$ 
GeV in Eq.~(\ref{width}).

Before presenting our results we first explain the numerical procedure 
used for our calculations. Because the differential dilepton rate 
$dR^{l^+l^-}/d^4P$ given in Eq.~(\ref{scattering}) is independent of 
the assumed space-time model, we first evaluate it numerically using 
double-exponential integration with a target precision of $10^{-9}$. 
The result for the rate was then tabulated on a uniformly-spaced 
4-dimensional grid in $M$, $p_T$, $y$, and $\xi$: $M/p_{\rm hard}, 
p_T/p_{\rm hard} \in \{0.1,20\}$, $y \in \{-10,10\}$ and $\xi \in 
\{0,5\}$.  This table was then used to build a four-dimensional 
interpolating function which was valid at continuous values of these 
four variables. We then boost this rate from the local reference frame 
to center-of-mass frame and evaluate the remaining integrations over 
space-time ($\tau$ and $\eta$), transverse momentum and invariant mass 
appearing in Eq.~(\ref{yspectrum}) using quasi-Monte Carlo integration 
with $\tau \in \{\tau_0,\tau_f\} $, $\eta \in \{-10,10\}$ and, 
depending on the case, restrict the integration to any cuts specified 
in $M$ or $p_T$. 

Our final integration time, $\tau_f$, is set by solving numerically 
for the point in time at which the temperature in our interpolating 
model is equal to the critical temperature, i.e. $p_{\rm 
hard}(\tau_f,\eta) = T_C$. We will assume that when the system reaches 
$T_C$, all medium emission stops. Note that due to the fact that 
$p_{\rm hard}$ depends on the parton rapidity, the plasma lifetime now 
depends on which rapidity slice you are in, with higher rapidities 
having a shorter lifetime due to their lower initial ``temperature''. 
We are not taking into account the emission from the mixed/hadronic 
phase at late times since the kinematic regime we study (high $M$ and 
$p_T$) is dominated by early-time high-energy dilepton emission 
\cite{Mauricio:2007vz,Strickland:1994rf}.

\begin{figure*}[t]
\begin{center}
 \includegraphics[width=11cm]{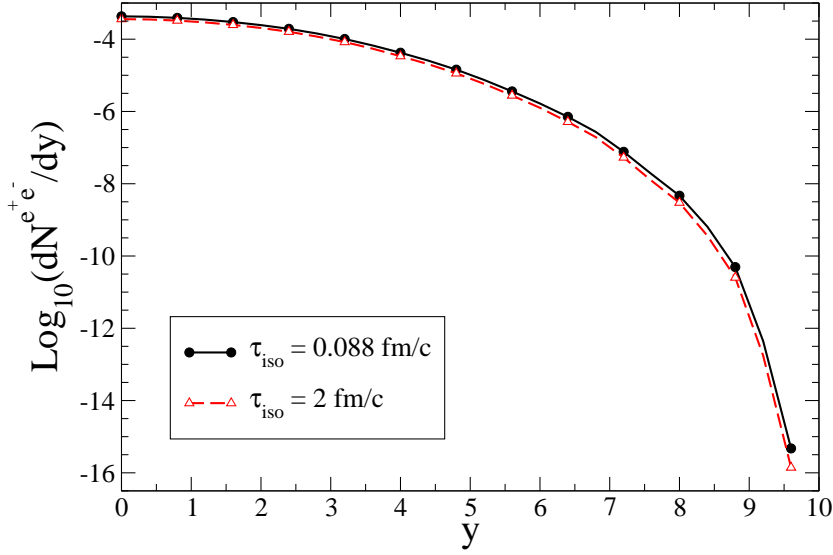}
\end{center}
\vspace{-2mm}
\caption{Fixed final multiplicity condition collisionally-broadened 
interpolating model dilepton yields as a function of rapidity in Pb-Pb 
collisions at LHC, with a cut $M \geq 2$ GeV and $P_T \geq 100$ MeV. For 
medium dileptons we use $\gamma$ = 2 and $\tau_{\rm iso}$ to be either 
0.088 or 2 fm/c for LHC energies.}
\label{fig:4}
\end{figure*}

We show our predicted dilepton spectrum as a function of the pair 
rapidity, $y$, for LHC energies using our model described by 
Eqs.~(\ref{eq:modelEQs}) in Fig.~\ref{fig:4}. From this, we see that 
for LHC energies there is a suppression when we vary the isotropization 
time from $\tau_0$ to 2 fm/c. This suppression can be explained 
qualitatively by two mechanisms. The first one, the anisotropic nature 
of the distribution function as a consequence of the rapid expansion 
implies that dileptons with larger values of longitudinal momentum are 
reduced compared with the case of an isotropic distribution function. 
The suppression will depend on the maximum amount of momentum-space 
anisotropy achieved at early times and also on the time dependence of 
the anisotropy parameter $\xi$; in this work, we consider a realistic 
scenario for a collisionally-broadened plasma. The other source of 
rapidity dependence of the final dilepton spectra is related to the 
fact that the hard momentum scale (``temperature'') depends explicitly 
on the rapidity $\eta$, even in the case of instantaneous 
thermalization. To generate Fig.~\ref{fig:4} we have applied a cut $M 
\geq$ 2 GeV and $P_T \geq 100$ MeV.  As can be seen from this figure a 
isotropization time of $\tau_{\rm iso}=$ 2 fm/c results in fewer 
dileptons as compared to ``instantaneous'' isotropization $\tau_{\rm 
iso}=$ 0.088 fm/c.  This suppression is enhanced at forward rapidities.

%%%%%%%%%%%%%%%%%%%%%%%%%%%%%%%%%%%%%%%%%%%%%%%%%%%%%%%%%%%%%%%%%%%%%%
%%%%%%%%%%%%%%%%%%%%%%%%%%%%%%%%%%%%%%%%%%%%%%%%%%%%%%%%%%%%%%%%%%%%%%

In order to quantify the effect of the pre-equilibrium emission we 
define the ``dilepton modification'' factor as the ratio of the dilepton 
yield obtained with an isotropization time of $\tau_{\rm iso}$ to that 
obtained from an instantaneously thermalized plasma undergoing only 
1+1 hydrodynamical expansion, ie. $\tau_{\rm iso}=\tau_0$:
\begin{equation} 
\Phi(\tau_{\rm iso}) \equiv \left. 
\left( \dfrac{dN^{e^+e^-}(\tau_{\rm iso})}{dy} \right) \right/ \left( 
\dfrac{dN^{e^+e^-}(\tau_{\rm iso}=\tau_0)}{dy} \right) \; .
\end{equation} 

This ratio measures how large the effect of early-time momentum 
anisotropies are on medium dilepton production. In the case of 
instantaneous isotropization, $\Phi(\tau_{\rm iso})$ is unity, and for 
$\tau_{\rm iso} > \tau_0$ any deviation from unity indicates a 
modification of medium dilepton production due to pre-equilibrium 
emissions. 

%%%%%%%%%%%%%%%%%%%%%%%%%%%%%%%%%%%%%%%%%%%%%%%%%%%%%%%%%%%%%%%%%%%%%%%%%%%%%%
\begin{figure*}[t]
\begin{center}
 \includegraphics[width=11cm]{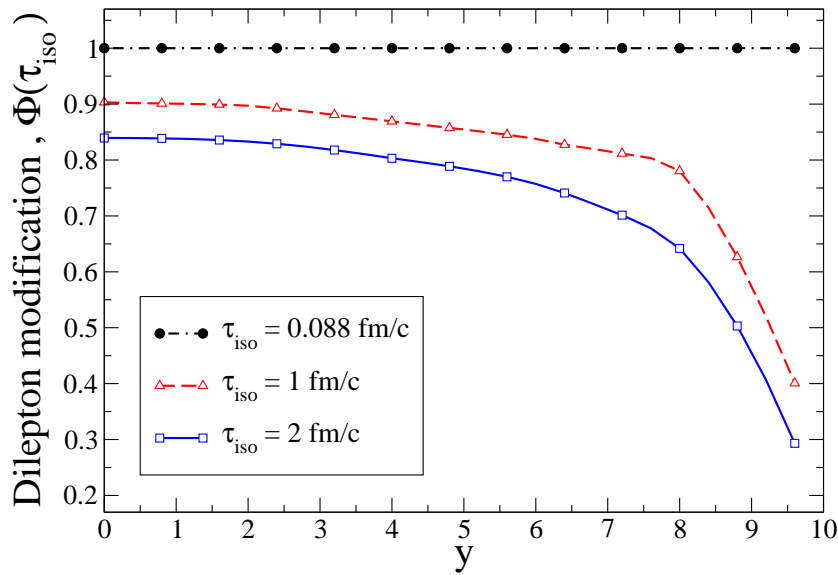}
\end{center}
%\vspace{-2mm}
\caption{Predicted dilepton modification factor,  $\Phi(\tau_{\rm 
iso})$, for three different assumed plasma isotropization times, 
$\tau_{\rm iso} \in \{$ 0.1,1,2$\}$ fm/c.  Cuts are the same as in 
Fig.~\ref{fig:4}.}
\label{fig:5}
\end{figure*}
%%%%%%%%%%%%%%%%%%%%%%%%%%%%%%%%%%%%%%%%%%%%%%%%%%%%%%%%%%%%%%%%%%%%%%%%%%%%%%

In Fig.~\ref{fig:5} we show our prediction for the rapidity dependence 
of the high-energy dilepton modification factor, $\Phi(\tau_{\rm iso})$, 
for three different assumed plasma isotropization times, $\tau_{\rm 
iso} \in \{$ 0.1,1,2$\}$ fm/c.  To generate this figure we have 
applied a cut $M \geq$ 2 GeV and $P_T \geq 100$ MeV.  As can be seen 
from this figure a isotropization time of $\tau_{\rm iso}=$ 2 fm/c 
results in fewer dileptons as compared to ``instantaneous'' 
isotropization $\tau_{\rm iso}=$ 0.088 fm/c.  This suppression is 
enhanced at forward rapidities and reaches a maximum suppression of a 
factor of 3 at extremely forward rapidities.

Using the dilepton modification factor as our criterion we find that 
for our collisionally-broadened interpolating model with fixed final 
multiplicity, the dilepton yields as a function of the rapidity at 
$\tau_{iso}$ = 2 fm/c can be suppressed up to $\sim$ 20\% for 0$< y 
\lesssim$ 4. The suppression of dilepton yields is more dramatic at 
rapidity values around $y \sim$ 9 and can be as large as a factor of 3. 
With sufficiently accurate experimental results this could give an 
experimental method for determining the isotropization time of a quark 
gluon plasma as formed in an ultrarelativistic nuclear collision.

%%%%%%%%%%%%%%%%%%%%%%%%%%%%%%%%%%%%%%%%%%%%%%%%%%%%%%%%%%%%%%%%%%%%%%%%%%%%%%%%%%%%%%%%%%%%%%%%%%%%%%%%%%%%%%%%%%%%%%%%%%%%%%%%%%%%%%%%%%%%%%%%%%%%%%%%%%%%%%%%%%%%%%%%%%%%%%%%%%%%%%%%%%%%%%%%%%%%%%%%%%%%%%%%%%%%%%%%%%

\section{Conclusions}
\label{sec:6}

In this work we have introduced a phenomenological model that takes 
into account early-time momentum-space anisotropies in the rapidity 
dependence of high-energy dilepton production. To do this we have 
modeled the temporal evolution of the plasma anisotropy parameter 
$\xi$ and the hard momentum scale $p_{\rm hard}$.  To study the 
dilepton production rapidity dependence, we have parametrized the 
rapidity dependence of $p_{\rm hard}$ using a Gaussian profile which 
is consistent with experimental observations of final pion spectra 
from AGS through RHIC energies.

We have applied the proposed model to study high-energy dilepton 
yields as a function of the pair rapidity and find that this observable is 
sensitive to the chosen value of $\tau_{\rm iso}$. This suppression 
can be explained as a consequence of the combined effect of the 
anisotropy in momentum-space achieved at early-times due to expansion 
and the rapidity dependence of the hard momentum scale which explicitly 
breaks longitudinal boost invariance. We find that with the resulting 
dilepton modification factor, $\Phi(\tau_{\rm iso}$=2 fm/c), shows 
suppressed dilepton yields in the forward rapidity region which can be 
up to 20\% for 0 $< y \lesssim$ 4 and  up to a factor of 3 at $y \sim$ 
9. The amplitude of the suppression of $\Phi(\tau_{\rm iso})$ could 
help us to experimentally constrain $\tau_{\rm iso}$ given 
sufficiently precise data in the forthcoming LHC experiments. In this 
way forward dileptons would provide a way to determine the plasma 
isotropization time experimentally.

An uncertainty of our treatment comes from our implicit assumption of 
chemical equilibrium. If the system is not in chemical equilibrium 
(too many gluons and/or too few quarks) early time quark chemical 
potentials, or fugacities, will affect the production of lepton pairs 
\cite{Dumitru:1993vz,Strickland:1994rf}.  However, to leading order, 
the quark and gluon fugacities will cancel between numerator and 
denominator in the dilepton suppression factor, $\Phi(\tau_{\rm iso})$ 
\cite{Strickland:1994rf}.  We, therefore, expect that to good 
approximation one can factorize the effects of momentum space 
anisotropies and chemical non-equilibrium.

We note in closing that the interpolating model presented here can be 
applied to other observables than dilepton yields. Indeed, with this 
model it is possible to assess the phenomenological consequences of 
momentum-space anisotropies on other observables which are sensitive 
to early-time stages of the QGP, e.g. photon production, heavy-quark 
transport, jet-medium induced electromagnetic and gluonic radiation, 
etc.
 
%%%%%%%%%%%%%%%%%%%%%%%%%%%%%%%%%%%%%%%%%%%%%%%%%%%%%%%%%%%%%%%%%%%%%%

\section*{Acknowledgments}

We thank A. Dumitru, S. Jeon, M. Bleicher, H. Appelsh\"auser and B. 
Schenke for helpful discussions. M. Martinez thanks N. Armesto and C. 
Salgado for assistance provided in order to attend the Hard Probes 
2008 conference where this work was initiated. M. Martinez was 
supported by the Helmholtz Research School and Otto Stern School of 
the Goethe-Universit\"at Frankfurt am Main. M.S. was supported by DFG 
project GR 1536/6-1. M.S. also acknowledges support from the Yukawa 
Institute for Theoretical Physics during the ``Entropy Production 
Before QGP'' workshop.

%%%%%%%%%%%%%%%%%%%%%%%%%%%%%%%%%%%%%%%%%%%%%%%%%%%%%%%%%%%%%%%%%%%%%%

\end{document}